\documentclass[prl,twocolumn,superscriptaddress,nopacs]{revtex4}
\usepackage{times}
\usepackage{amsmath}
\usepackage{amsthm}
\usepackage{amssymb}
\usepackage{amsbsy}
\usepackage{amsfonts}
\usepackage{graphicx}
\usepackage{subfigure}
\usepackage{textcomp}
\usepackage[mathscr]{euscript}
\usepackage[english]{babel}
\usepackage{bibunits}
\begin{document}
\title{Magnetic field-tuned Aharonov--Bohm oscillations and evidence for
non-Abelian anyons at $\nu=5/2$}

\author{R.~L.~Willett}
\affiliation{Bell Laboratories, Alcatel-Lucent, Murray Hill, New Jersey 07974,
USA}
\author{C.~Nayak}
\affiliation{Microsoft Research, Station Q, Elings Hall, University of
California, Santa Barbara, California 93106, USA}
\affiliation{Department of Physics, University of California, Santa Barbara,
California 93106, USA}
\author{K.~Shtengel}
\affiliation{Microsoft Research, Station Q, Elings Hall, University of
California, Santa Barbara, California 93106, USA}
\affiliation{Department of Physics and Astronomy, University of California,
Riverside, California 92521, USA}
\author{L.~N.~Pfeiffer}
\affiliation{Department of Electrical Engineering, Princeton University,
Princeton, New Jersey 08544, USA}
\author{K.~W.~West}
\affiliation{Department of Electrical Engineering, Princeton University,
Princeton, New Jersey 08544, USA}

\begin{abstract}
We show that the resistance of the $\nu=5/2$
quantum Hall state, confined to an interferometer, oscillates
with magnetic field consistent with an Ising-type non-Abelian state.
In three quantum Hall interferometers of different sizes, resistance
oscillations at $\nu=7/3$ and integer filling factors have the magnetic field period expected
if the number of quasiparticles contained within
the interferometer changes so as to keep the area and the total charge
within the interferometer constant. Under these conditions,
an Abelian state such as the $(3,3,1)$ state would show
oscillations with the same period as at an integer quantum Hall state.
However, in an Ising-type non-Abelian state there would be a rapid oscillation
associated with the {\textquotedblleft}even-odd effect{\textquotedblright} and a
slower one associated with the accumulated Abelian phase due to both the
Aharonov-Bohm effect and the Abelian part of the quasiparticle braiding
statistics. Our measurements at $\nu=5/2$ are consistent with the latter.
\end{abstract}
\date{\today}
\maketitle
\begin{bibunit}[apsrev]
\paragraph{Introduction.}
The origin of the fractional quantum Hall effect~\cite{Tsui1982} at
filling factor ${\nu}=5/2$~\cite{Willett1987,Pan1999b,Eisenstein2002}
has been a long-standing open issue, which is important
because it has been conjectured that this state of matter supports non-Abelian anyons
~\cite{Moore1991,Greiter1992,Nayak1996c,Read2000,Bonderson2010c}.
Two point-contact Fabry--P\'{e}rot interferometers have been proposed
to observe the Aharonov--Bohm (AB) effect and the anyonic braiding statistics of quasiparticles~\cite{Chamon1997}. In a non-Abelian state, not only the phase but the also the amplitude of the observed oscillations is indicative of the braiding statistics~\cite{Fradkin1998,DasSarma2005,Stern2006a,Bonderson2006a}.  Specifically, if the ${\nu}=5/2$ state is indeed non-Abelian, the quasiparticle parity within the interferometer dictates whether the resistance of an interferometer oscillates with enclosed area (controlled by a side gate) with a period associated with charge $e/4$ quasiparticles. Such oscillations should only be seen when the parity is even -- the {\textquotedblleft}even-odd effect{\textquotedblright}~\cite{Stern2006a,Bonderson2006a}. Previous experiments~\cite{Willett2009a,Willett2010a}
are broadly consistent with these predictions~\cite{Wan2008,Bishara2009a,Bishara2009b,Rosenow2009a}, although some puzzles
remain, as we discuss below.
%(For both parities, oscillations with a period associated with Abelian quasiparticles of charge $e/2$ should be observed, but in the limit of weak backscattering and low temperature they should have smaller amplitude.) Previous experiments~\cite{Willett2009a,Willett2010a} have observed $e/4$ oscillations
%at $\nu=5/2$. As the side-gate voltage is swept, these oscillations disappear
%and reappear, consistent with alternating parity of the enclosed $e/4$ quasiparticles~\cite{Wan2008,Bishara2009a,Bishara2009b,Rosenow2009a}. When $e/4$ oscillations are not present, $e/2$ oscillations are visible, yet it is not clear whether the two coexist, as theory requires. By varying the magnetic field by an amount discussed below, the intervals of side-gate voltage with present/absent $e/4$ oscillations are interchanged~\cite{Willett2010a}.

\begin{figure}[h]
  \begin{center}
    \includegraphics[width=1\columnwidth]{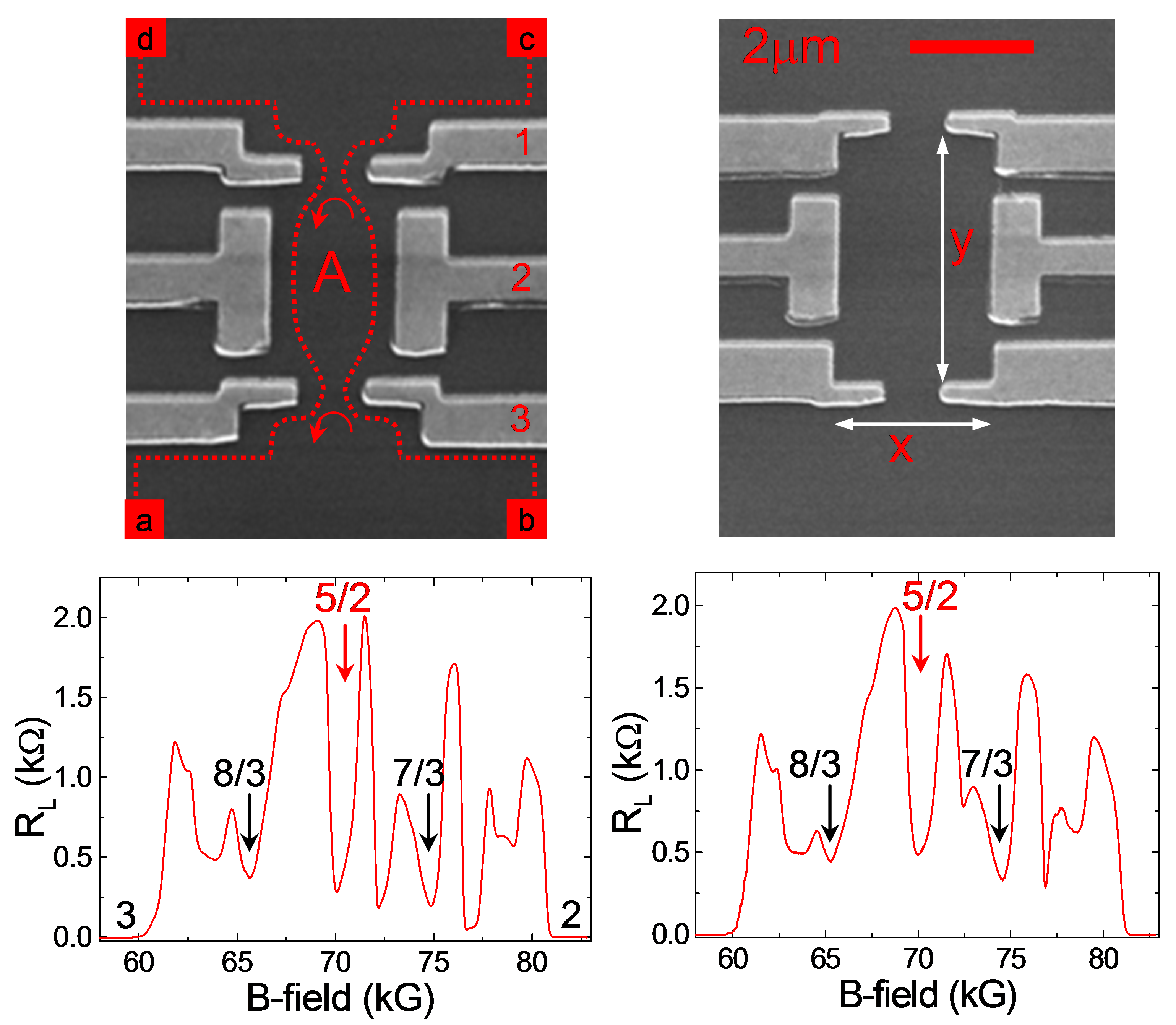}
  \end{center}
  \caption[]{(top) Electron micrographs of two of the three interferometers used
in the measurements reported in this paper. The contacts are indicated schematically in the interferometer on the left by \textbf{a-d}. Current injected at contact \textbf{b} can be backscattered at the two
quantum point contacts shown, thereby defining an interference loop of area $A$.
(bottom) The longitudinal resistance $R_\text{L}$ for the two samples shows minima corresponding to
fractional quantum Hall states at $\nu=$\,7/3, 5/2, and 8/3.}
 \label{fig:devices}
\end{figure}

In this paper, we examine the magnetic field dependence of the resistance of a series of interferometers with a large range of active areas. Two of them are shown
in Fig.~\ref{fig:devices}. We formulate a model
%of the physics of the interferometer
based on the assumption that
the total charge in the interferometer and the enclosed area both
remain constant as the magnetic field is varied. We test it at $\nu=7/3$ and integer filling factors and show that it is consistent with
the experimental data -- in the $\nu=7/3$ case, it predicts a resistance oscillation
with the somewhat surprising flux period ${\Phi_0}/2$. We thereby determine
the effective area of the interference loop in each device
(and each `preparation' of each device, which we describe in the next paragraph).
The model also predicts that the resistance in the $\nu=5/2$ state
will oscillate as the product of two oscillations, one with flux period
${\Phi_0}/5$ and the other with flux period ${\Phi_0}$, as we explain
and compare to our experimental data below.

\paragraph{Interferometers.}
The interferometers used in this paper are fabricated
from high-mobility ($28\times 10^6$\,cm$^2$/V$\cdot$s),
high-density ($4.2 \times 10^{11}$\,cm$^{-2}$) GaAs/AlGaAs quantum well
heterostructures. A $40$\,nm SiN layer is applied to the heterostructure.
The size and shape of the 2D electron channel, which is $200$ nm below the surface,
is controlled by $100$ nm thick Al top gates that are deposited on the SiN layer,
as shown in Fig.~\ref{fig:devices}. Prior to charging the top gates, the samples can be briefly illuminated to enhance mobility and to provide different sample preparations since the illumination changes the distribution of localized charges in the device,
as do different cool-downs~\cite{Willett2009a,Willett2010a,Willett2007}.
Further description of the device preparations and measurement details is presented in Supplemental Material. Two interfering edge currents result from quasiparticle tunneling across constrictions defined by gate sets 1 and 3. The longitudinal resistance $R_\text{L}$ is measured with contacts labeled \textbf{a} through \textbf{d} in the electron-micrograph in Fig.~\ref{fig:devices}(a) by the voltage drop from contact \textbf{a} to \textbf{d}, with current driven from \textbf{b} to \textbf{c}, using standard lock-in techniques. The two standard top gate designs shown in electron micrographs in Fig.~\ref{fig:devices}(a) are labeled with device dimension parameters \textit{x} and \textit{y} adjusted to produce three separate samples with ratios of areas of roughly 3:2:1. In addition, the functional areas in these devices are defined by applying the gate voltages, resulting in a range of areas from 0.1 to 0.6\,{\textmu}m$^2$. The temperature of the measurements is ~20mK in all data presented here.

\paragraph{Previous Results.}
Oscillations in $R_\text{L}$ have previously been observed as a function of side gate voltage, $V_\text{s}$, which
changes the area of the interferometer~\cite{Willett2009a,Willett2010a}. Interpreted
as due to the Aharonov-Bohm effect, the expected period of oscillation is
${\Delta} V_\text{s} \propto \Delta A = (e/e^*) {\Phi_0}/B$
($\Phi_0=hc/e$ is the fundamental flux quantum and
$e^*$ is the charge of the interfering quasiparticles), from
which the quasiparticle charge $e^*$ could be obtained if the proportionality
constant between ${\Delta} V_\text{s}$ and $\Delta A$ were known.
Assuming its independence of magnetic field, this constant could be determined from the period of
$R_\text{L}$ oscillations at integer filling factors, where ${e^*}=e$,
or at $\nu=7/3$, where ${e^*}=e/3$
is expected. Both filling fractions give similar proportionality constants
between ${\Delta} V_\text{s}$ and $\Delta A$, which supports the idea
that this constant is approximately independent of magnetic field. At $\nu=5/2$,
$R_\text{L}$ oscillations  in some intervals of
$V_\text{s}$ appear consistent with AB oscillations corresponding to
$e/4$ charges while in other intervals they seem consistent with $e/2$
charges~\cite{Willett2009a,Willett2010a}. These
results have been interpreted as a manifestation of the
``even-odd effect'' \cite{Bonderson2006a,Stern2006a}:
charge $e/4$ oscillations should be observed only when
there is an even number of charge $e/4$ quasiparticles in the interferometer;
charge $e/2$ oscillations should always be observed. When there
is an odd number of charge $e/4$ quasiparticles in the interferometer,
this should be the only type of oscillation visible \cite{Bishara2009a}.
In Refs.~\onlinecite{Willett2009a,Willett2010a} only e/2 oscillations are visible in certain side-gate voltage intervals (when, according to this interpretation, an odd number of e/4 quasiparticles is in the interferometer), but it is not clear whether both e/4 and e/2 oscillations -- or only e/4 -- are present in the other intervals.

By contrast, in this letter we focus on $R_\text{L}$
measurements during magnetic field sweeps. At integer filling,
a $B$-field sweep produces AB oscillations of $R_\text{L}$ with
period ${\Delta}B\cdot A={\Phi_0}\approx {41}$\,G\,{\textmu}m$^2$,
where $A$ is the current-encircled area of the interferometer; we thereby
determine the active area for each of the different
devices and sample preparations.

\paragraph{Model.}
The key assumption in our interpretation of the experimental data
is that the charge contained within the interference loop
and the area of the loop remain constant as the magnetic field is varied.
It is natural to assume that the charge contained within the loop remains constant
if it is primarily determined by the local electrostatic potential or, in other words,
if the Coulomb energy dominates. In such a case, as the magnetic field is varied,
one of two possibilities will occur. Quasiparticles will be created in the bulk
or else the quantum Hall droplet will shrink or expand; in the former case,
the area of the interference loop will remain constant.
We expect this scenario to hold if there are localized states in the bulk
that have very low energy as a result of disorder so that it is energetically
favorable to create quasiparticles there, rather than to change the charge density
at the edge. When this scenario holds, increasing the flux through the interferometer
by $\Phi$ causes the number of charge $e^*$ quasiparticles to change by
$N_{e^*} = (\nu\Phi/\Phi_0)/({e^\ast}/e)$.

Meanwhile, changing the flux by $\Phi$ and the number of
charge $e^*$ quasiparticles by $N_{e^*}$ causes a change
${\Delta}{\gamma}$ in the phase acquired by a quasiparticle taking
one path around the interferometer relative to the phase
acquired by a quasiparticle going around the other:
\begin{eqnarray}
{\Delta}{\gamma} &=& 2 {\pi} ({\Phi}/{\Phi}_0) (e^\ast/e) - 2{\theta_{e^*}} N_{e^*}\cr
&=& ({\Phi}/{\Phi}_0)[2 {\pi}(e^\ast/e) - 2{\theta_{e^*}}  (\nu e/e^\ast)]
\label{eq:AB_phase_general}
\end{eqnarray}
The first term on the right-hand-side is the (ordinary electromagnetic) AB phase seen by a
charge $e^\ast$ quasiparticle encircling flux ${\Phi}$.
The second term is the statistical phase seen by a charge $e^\ast$ quasiparticle when it
encircles $N_{e^*}$ such quasiparticles; the phase acquired when a
single charge $e^\ast$ quasiparticle encircles another is $2{\theta_{e^*}}$,
assuming that the particles are Abelian.
For non-Abelian particles, more care is required, as we will see below.
%In going from the first line in Eq. (\ref{eq:AB_phase_general})
%to the second, we have used $N_{e^*} = (\nu\Phi/\Phi_0)/({e^\ast}/e)$.
The relative minus sign can be understood using the argument in~\cite{Chamon1997}, where it is explained why the AB and statistical phases should cancel under certain conditions.

In an integer quantum Hall state, ${e^*}=e$ and $\theta_{e}=\pi$,
so ${\Delta}{\gamma}=2\pi({\Phi}/{\Phi}_0)$ and $R_\text{L}$ will
oscillate with magnetic field period
$\Delta B_{0} \cdot A = \Phi_0 \approx 41$\,G\,{\textmu}m$^2$. Now consider the
$\nu=7/3$ state. If it is in the same universality class as the $\nu=1/3$ Laughlin state,
then ${e^*}=e/3$ and $2\theta_{e/3}=2\pi/3$. Then
${\Delta}{\gamma}=-4\pi({\Phi}/{\Phi}_0)$. Consequently, $R_\text{L}$ will
show oscillations with period $\Delta B_{1} \cdot A = {\Phi_0}/2 \approx 20$\,G\,{\textmu}m$^2$
-- i.e half that in an integer quantum Hall state.

Now consider the case of $\nu=5/2$. If the system is in an Ising-type
topological phase such as the Moore--Read state~\cite{Moore1991} or the anti-Pfaffian
state~\cite{Levin2007a,Lee2007a}, then when there is an even number of charge $e/4$ quasiparticles in the
interference loop, the Ising topological charge will be $1$ or $\psi$,
but when there is an odd number in the interference loop,
the Ising topological charge will be $\sigma$. As a result, if
one particular topological charge is energetically favorable
for even quasiparticle number -- let us suppose, for the sake of
concreteness, that it is $1$ -- then the non-Abelian Ising topological charge has
a periodicity of two quasiparticles or, taking
$N_{e/4} = (\nu e \Phi/\Phi_0)/(e/4)$, a flux period
$\Phi = 2{\Phi_0}(e/4e)/\nu={\Phi_0}/5$. Hence,
$R_\text{L}$ oscillates with magnetic field period
$\Delta B_2 \cdot A \approx 8$\,G\,{\textmu}m$^2$.
However, there is also an Abelian phase
(\ref{eq:AB_phase_general}) which can have a different periodicity.
The Abelian phase acquired when a charge $e/4$ quasiparticle encircles
an $2N$ quasiparticles with Ising charge $1$ (or, equivalently, $N$
charge $e/2$ quasiparticles) is $\theta=\pi/4$ and $N=(\nu e \Phi/\Phi_0)/(e/2)$.
Hence, Eq.~(\ref{eq:AB_phase_general})
now reads: $\Delta \gamma = -2\pi (\Phi/\Phi_0)$. Therefore, there is also a slower
oscillation in $R_\text{L}$ with magnetic field period
$\Delta B_0 \cdot A = \Phi_0 \approx 41$\,G\,{\textmu}m$^2$.

If, however, the Ising charge is not fixed to $1$
for any even number of quasiparticles, but may be randomly either $1$ or $\psi$, then
the slower, period $\Phi_0$, oscillation will be afflicted by
random $\pi$ phase shifts that could wash it out.
If the system were in an Abelian $(3,3,1)$ state,
then similar considerations lead to a period $\Phi_0$ oscillation
but no rapid period ${\Phi_0}/5$ oscillation.

\begin{figure}[t!]
  \begin{center}
    \includegraphics[width=1\columnwidth]{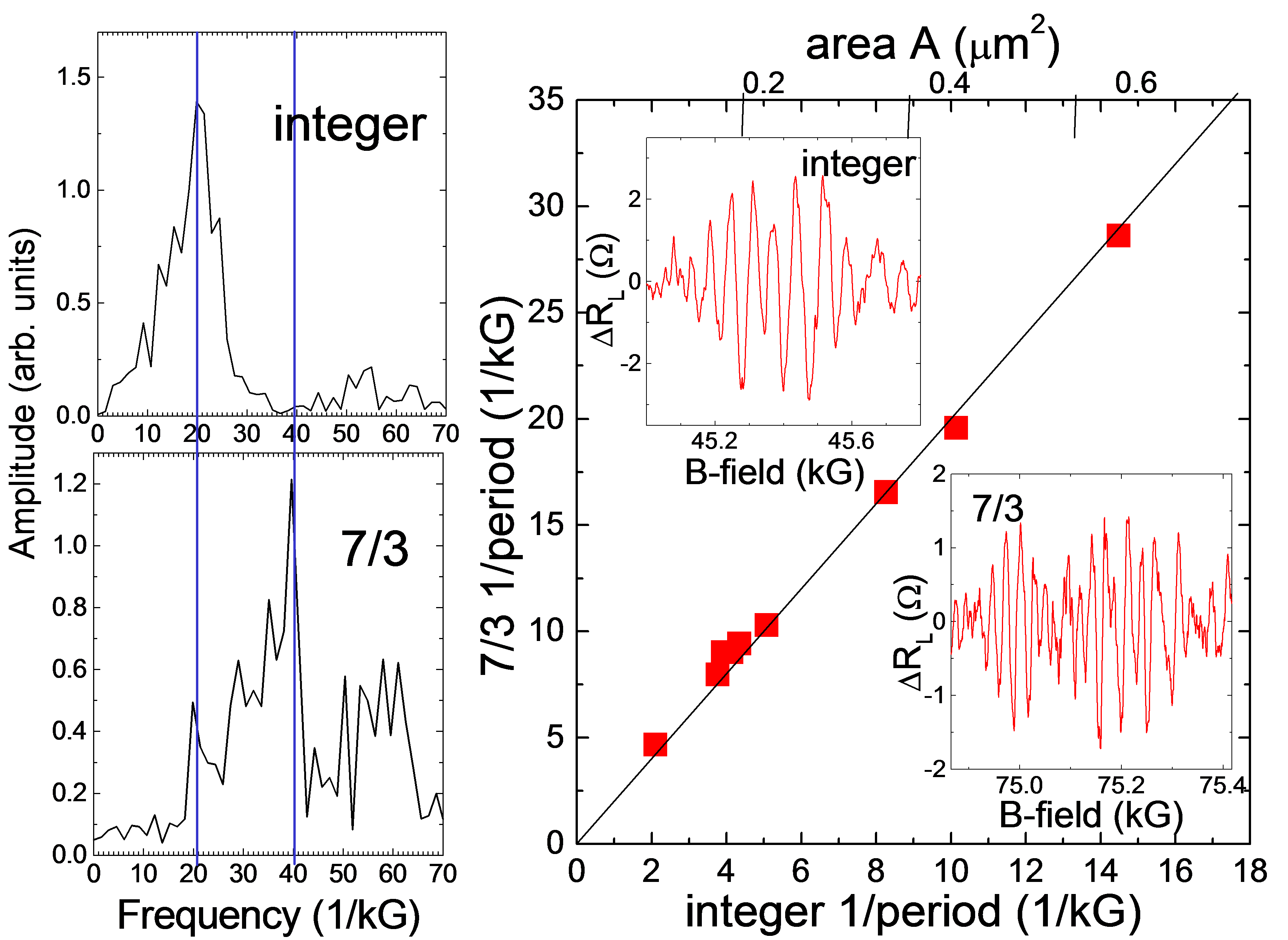}
  \end{center}
  \caption[]{Oscillations with magnetic field at the integer state $\nu=4$ and also at $\nu=7/3$ (insets), with FFTs of these oscillations shown in the left hand panels. The ratio between
these two oscillation periods is the same in eight device preparations of varying size.}
  \label{fig1}
%  \label{fig:devices}
\end{figure}
\paragraph{Comparison with experiment.}
The overall $B$-sweep trace between filling factors 2 to 3
of $R_\text{L}$ across two of the interferometers is shown in the bottom two panels of
Fig.~\ref{fig:devices}; they clearly demonstrate fractional quantum Hall states at
$\nu=$\,7/3,~8/3, and 5/2. This overall trace is averaged locally to define a background
which we subtract from the raw $R_\text{L}$ measurement to make the
oscillations clearer. Measurement of these oscillation sets was repeated for multiple interferometric areas. The $\Delta B_0$ period should change with area according to our model. From the three devices used and the multiple preparations and gate values employed,
the measured $\Delta B_0$ periods show active areas ranging from 0.1\,{\textmu}m$^2$ to $\sim 0.6$\,{\textmu}m$^2$.

To put our picture to test, we first consider $\nu=4$ and $7/3$.
Oscillations of $R_\text{L}$ with $B$ are shown in the upper left
of Fig.~\ref{fig1}. The period $\Delta B_0$ of oscillations
is found to be similar near integer filling 2, 3 and 4 for each device,
consistent with this being an AB oscillation and not Coulomb effects~\cite{Willett2013a}.
From the periodicity $\Delta B_0$, we determine the active area of this preparation.
(For instance, for the preparation displayed in Fig.~\ref{fig3},
$\Delta B_0 \approx 110$~G, from which we deduce $A \approx 0.36$\,{\textmu}m$^2$.)

We now turn to $\nu=7/3$. The results for $B$-sweeps are shown in
Fig.~\ref{fig1}. Oscillations at 7/3
and integer filling factors are shown in Fig.~\ref{fig1} insets with their corresponding Fourier transforms, which show peaks. The 7/3 peak frequency is twice the $\nu=4$ peak
frequency (or half the period), consistent with the analysis above.
% that determined that the 7/3 $R_\text{L}$ oscillation period should be half that of the integer filling period.
The same $R_\text{L}$ measurements comparing $\nu=7/3$ and integer filling factors were carried out on the three different devices and different preparations of this study, as summarized in the right panel of Fig.~\ref{fig1}. The $R_\text{L}$ oscillations at $\nu=7/3$ consistently occur at twice the frequency of their respective integer filling factor oscillations over the full range of device areas studied. We conclude that the assumptions
and analysis outlined above are valid.

\begin{figure}[t!]
  \begin{center}
\includegraphics[width=0.8\columnwidth]{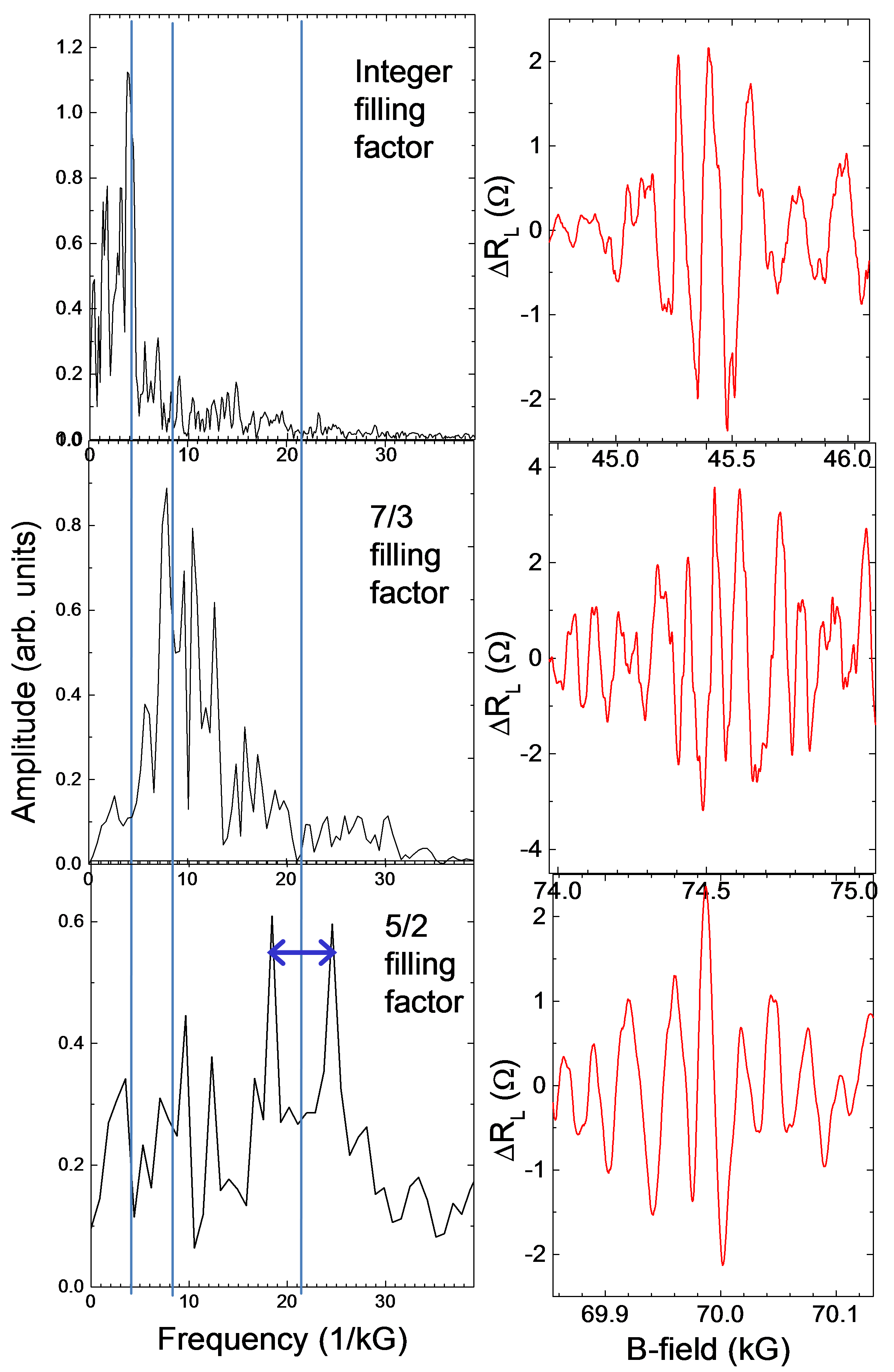}
 \end{center}
  \caption[]{Oscillations in $R_{\text{L}}$ as a function of magnetic field
and the associated Fourier transforms at (a) $\nu=4$, (b) $\nu=7/3$,
and (c) $\nu=5/2$. The vertical blue lines mark 5x, 2x, and the integer frequency. The oscillations at $\nu=7/3$ are observed to have twice
the frequency of those at $\nu=4$, which is consistent with the theoretical
model explained in the text. The oscillations at $\nu=5/2$ show beating
between a fast oscillation with a period that is $1/5$ that at $\nu=4$ and a slow one with the same period as at $\nu=4$}.
  \label{fig2}
%  \label{fig:AB-oscillations}
\end{figure}

%We now turn our attention to $\nu=5/2$.
%Interference effects at $\nu=5/2$ are the focus of Fig.~\ref{fig2}: the small period oscillations corresponding to the expression/suppression of non-Abelian $e/4$ quasiparticle interference are demonstrated here.
Fig.~\ref{fig2} presents the comparison between interference oscillations of $\Delta R_\text{L}$  at $\nu=5/2$, 7/3, and $\nu=4$ observed in the same sample/prepapration. (The overall $B$-sweep trace of $R_\text{L}$ for this interferometer is shown in the bottom right panel of Fig.~\ref{fig:devices}.) Sets of oscillations are shown with their respective Fourier transforms. Once again, the $\nu=7/3$ oscillations are observed at half the magnetic field period of those
at $\nu=3$. Interestingly, the oscillations at 5/2 contain a higher frequency component, and the FFT spectrum demonstrates the predominant frequency is 5 times that of the integer oscillation frequency. This value is consistent with the expected oscillation frequency for expression/suppression of non-Abelian $e/4$ interference due to the changing number of quasiparticles with varying magnetic field. Moreover, the peak centered around five times the integer frequency is \emph{split}, with the splitting being roughly twice the frequency
observed at integer plateau. This corresponds to beats which are further consistent with the above prediction for the interplay between the AB and statistical contributions for a \emph{non-Abelian} $\nu=5/2$ state. Other such data sets are presented in supplemental materials.

\begin{figure}[t!]
  \begin{center}
    \includegraphics[width=1\columnwidth]{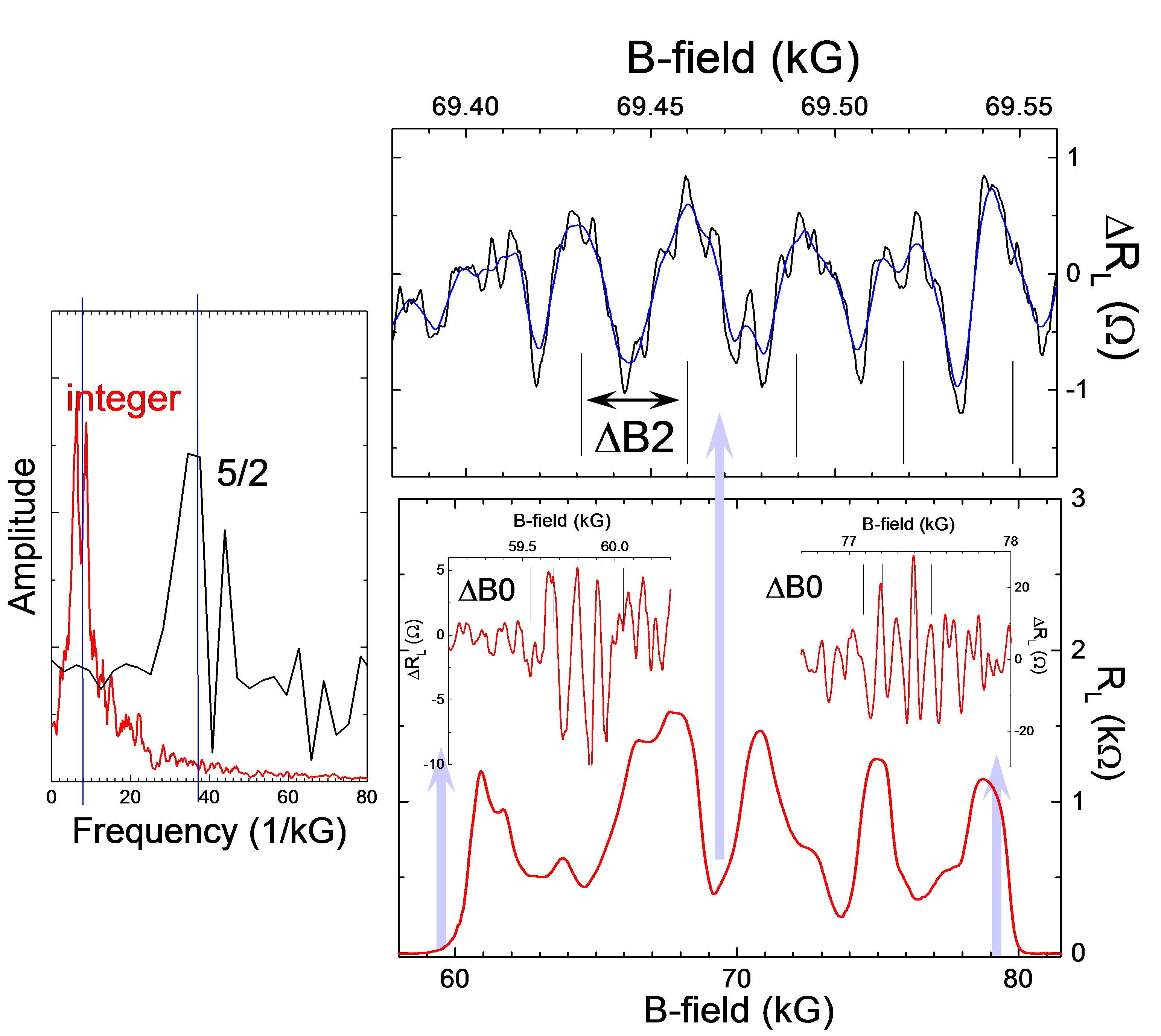}
  \end{center}
  \caption[]{Oscillations at $\nu=2$ (right inset), $\nu=3$ (left inset)and at $\nu=5/2$
(top panel). From the Fourier transforms
of $R_{\rm L}$ vs. $B$ (left panel), we see that the oscillation period at
$\nu=5/2$ is $1/5$ as large as at $\nu=integer$. Sometimes it shows beating
with a more rapid oscillation with the same period as at $\nu=2$.}
  \label{fig3}
%  \label{fig:different-devices}
\end{figure}
\begin{figure}[h]
  \begin{center}
    \includegraphics[width=0.8\columnwidth]{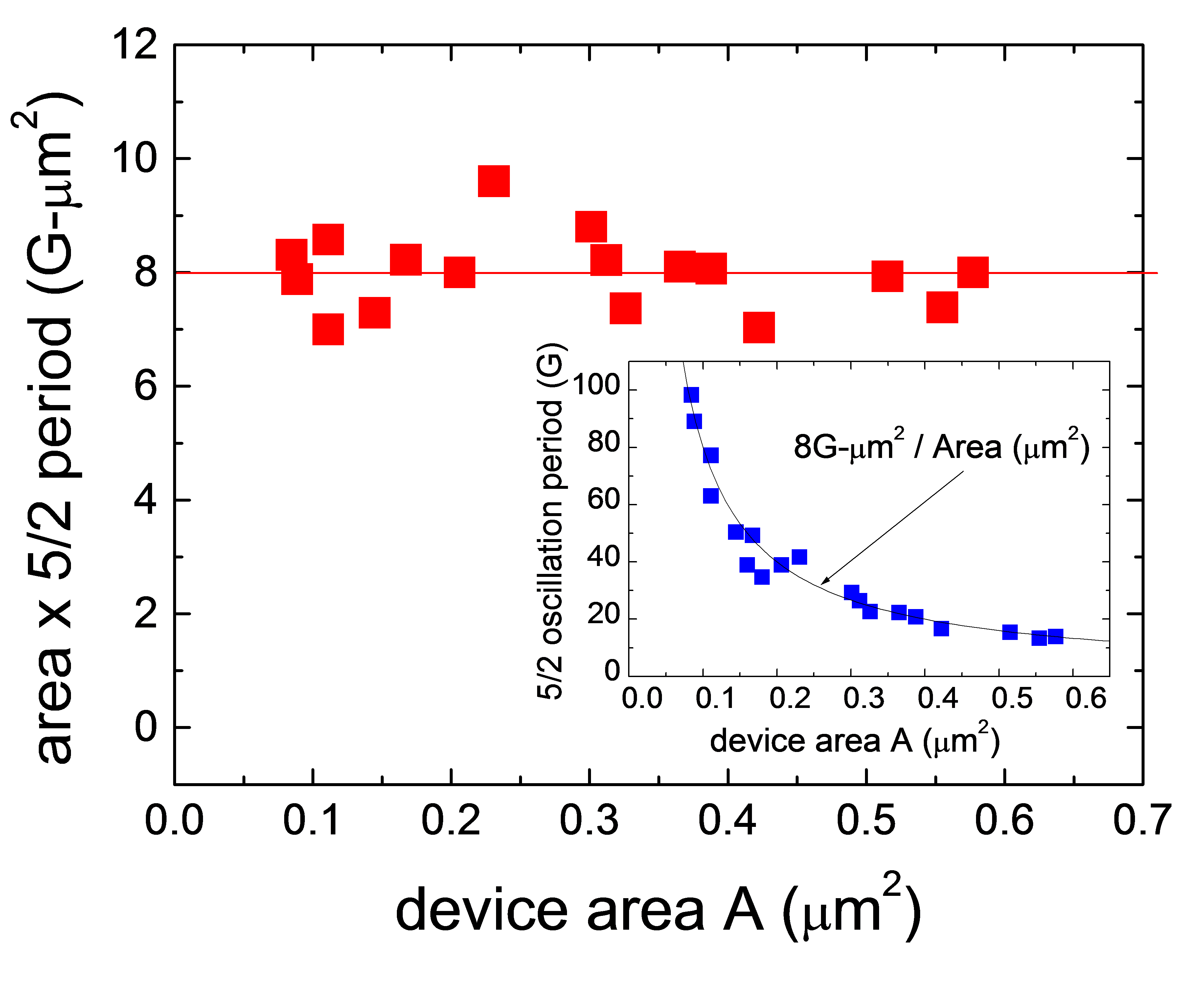}
  \end{center}
  \caption[]{The oscillation period in units of flux is independent of the
device area and is approximately $8$\,G\,{\textmu}m$^2 = \Phi_0/5$.
Equivalently (inset) the oscillation period in units of magnetic field is inversely proportional to the area.}
%\label{fig:AB-period}
  \label{fig4}
\end{figure}

The observation of a small oscillation period at $\nu=5/2$ in a series of samples with  different interferometer sizes and different sample preparations is further demonstrated in Figs.~\ref{fig3},~\ref{fig4}.  Fig.~\ref{fig3}, top panel, shows the $B$-sweep results for one of the sample preparations in device area 2, focusing on the small period resistive oscillations corresponding to multiple parity changes in the enclosed $e/4$ quasiparticle number near 5/2.
The set of oscillations is measured with no adjustments to the voltages of the quantum point contacts or central top gates. If our model is correct, the five periods of oscillation shown here represent ten parity changes. The $B$-field range near 5/2 where this data is taken is marked in the overall $R_\text{L}$ trace. The distinct resistive oscillations (black trace; the blue trace is a coarse smoothing of the data) near 5/2 in this preparation have a period ${\Delta}B_2\approx 22$\,G, compared to a period at integer fillings of ${\Delta}B_0\approx 110$\,G.
This is precisely the same fivefold ratio shown in  Fig.~\ref{fig2}, once again
in agreement with our model. Note here that splitting in the 5/2 peak is not resolved,
which may be an indication that the {\it fermion} parity in the interferometer
is not constant, a possibility discussed above. Or, alternatively, sweeps through
a wider $B$-field interval may be necessary to observe this slow oscillation
in some samples/preparations. Wider sweeps may also reveal
oscillations due to transport by charge-$e/2$ quasiparticles, which should have a period
$\Delta B \cdot A = {\Phi_0}/2$ (by essentially the same argument as for $7/3$).
They are not apparent in the magnetic field sweeps in Figs.~\ref{fig2}~,~\ref{fig3},
even though they are seen in side-gate voltage sweeps \cite{Willett2009a,Willett2010a}.

Fig.~\ref{fig4} summarizes the principal result of this study.  The oscillation period in units of flux (i.e., $\Delta B_2 \cdot A$) at $\nu=5/2$ measured for  different samples/preparations is approximately independent of
the device area derived from $\Delta B_0$. The observed values of $\Delta B_2 \cdot A$ is shown to be in reasonable agreement with the expected value of  8\,G\,{\textmu}m$^2$ which corresponds to the change in the parity of enclosed quasiparticles.

To conclude, this experiment provides the necessary complement to prior measurements~\cite{Willett2009a,Willett2010a} where AB oscillations were examined as a function of the active interferometer area $A$ controlled by the gate voltage. By sweeping the $B$ -field in these multiple area devices instead, the previous experimental limitation coming from slow gate charging has been avoided. The resistance oscillations observed near filling factor $\nu=5/2$ in multiple devises show a period consistent with the additional magnetic field needed to add one quasihole to their respective (different) active areas (thereby changing the quasiparticle parity). We stress that the presence of such a period is indicative of a non-Abelian nature of the $\nu=5/2$ state. While this interpretation is based on several assumptions discussed earlier, using the $\nu=7/3$ FQH state for control measurements significantly strengthens our case.
\begin{acknowledgments}
RLW, CN and KS acknowledge the hospitality of KITP supported in part by the NSF under grant PHY11-25915. CN and KS are supported in part by the DARPA-QuEST program. KS is supported in part by the NSF under grant DMR-0748925.
\end{acknowledgments}

%\bibliographystyle{apsrev}
%\bibliography{kirref}
%\putbib[kirref]

\end{bibunit}

\clearpage
%%%%%%%%%%%%%%%%%%%%%%%%%%%%%%%%%%%%%%%%%%%%%%%%%%%%%%%%%%%%%%%%%%%%%%%%%%%%%
%\renewcommand{\thesection}{S.\arabic{section}}
%\renewcommand{\thesubsection}{\thesection.\arabic{subsection}}
\setcounter{equation}{0}
\setcounter{figure}{0}
% Hack for making SOM Equations Conform to Science Format
%
% e.g. (S1), (S2), etc
% Requires AMS
\makeatletter %% With ams
\def\tagform@#1{\maketag@@@{(S\ignorespaces#1\unskip\@@italiccorr)}}
\makeatother
% Hack for making figures Say \figurename S\thefigure, e.g. Figure S1:
\makeatletter
\makeatletter \renewcommand{\fnum@figure}
{\figurename~S\thefigure}
\makeatother

% use bibnumfmt to change style at the end of the document
\renewcommand{\bibnumfmt}[1]{[S#1]}
% citenumfont command adds S to all numbers
\renewcommand{\citenumfont}[1]{S#1}
 
\renewcommand{\figurename}{Figure}
\begin{bibunit}[apsrev]
%%%%%%%%%%%%%%%%%%%%%%%%%%%%%%%%%%%%%%%%%%%%%%%%%%%%%%%%%%%%%%%%%%%%%%%%%%%%%

\section{Supplemental Materials}

\subsection{Theoretical results for $B$-sweep and $(3,3,1)$ state}
\label{sec:theoretical_results}

In this section we compare the expected oscillation
periods of magnetic field sweeps for Abelian and non-Abelian candidate states at $\nu=5/2$.
As explained in the main text, the net phase accumulation in a quantum Hall interferometer consists of two contributions: the AB phase due to the magnetic flux change by $\Phi$ and the statistical phase due to introducing $N_{e^*}$ additional quasiparticles of charge $e^*$  to the interferometer area:
\begin{equation}
{\Delta}{\gamma} = 2 {\pi} ({\Phi}/{\Phi}_0) (e^\ast/e) - 2{\theta_{e^*}}
N_{e^*}
\label{eq:net_phase_1}
\end{equation}
If we assume that the active area of the interferometer remains constant throughout a magnetic field sweep, these two contributions are not independent; $N_{e^*} = (\nu\Phi/\Phi_0)/({e^\ast}/e)$ and hence
\begin{equation}
{\Delta}{\gamma} = ({\Phi}/{\Phi}_0)\left[2 {\pi}\left(e^\ast/e\right) - 2{\theta_{e^*}}  \left(\nu e/e^\ast\right)\right].
\label{eq:net_phase_2}
\end{equation}
At $\nu=5/2$, the smallest charge excitation is an $e/4$ quasihole/quasipartice irrespective of the exact nature of the state. Since the smallest charge carriers are expected to dominate tunneling at the low temperature, weak tunneling limit~\cite{Fendley2007a,Bishara2009a}
(or, in the case of the anti-Pfaffian state, to vary with temperature in the same way
as transport due to $e/2$ quasiparticles),
we will first focus on these $e/4$ excitations.
As shown in the main text, if the $\nu=5/2$ state is non-Abelian
(whether it is Moore--Read or anti-Pfaffian), this should manifest
itself via a ${\Phi}_{0}/5$ oscillation period corresponding to the even-odd effect.
An additional ${\Phi}_{0}$ period of purely Abelian nature is also expected, although it may be washed out by fermion parity fluctuations.

If, on the other hand, the $\nu=5/2$ state is an Abelian  $(3,3,1)$ state~\cite{Halperin1983}, the ${\Phi}_{0}/5$ oscillation period should not be observed -- there is no even-odd effect in this case.
An Abelian phase can be calculated using Eq.~(\ref{eq:net_phase_2}) with a slight caveat.
The statistical angle
${\theta}$ can be either $3{\pi}/8$ or $-{\pi}/8$, depending on whether the spin of a
quasiparticle going around the interferometer and that of a quasiparticle inside the loop are the same or opposite. Since the $(3,3,1)$ state is spin-unpolarized, excitations of both spins may
carry charge around the interference loop. Hence we can use the average value
of the statistical angle, i.e. ${\pi}/8$, per quasiparticle (or, more precisely, ${\pi}/4$
per pair with opposite spins).  This yields ${\Delta}{\gamma} =2{\pi}
\left({\Phi}/{\Phi_0}\right)\left[(1/4)-(1/8)\times 10\right] = - 2{\pi}
\left({\Phi}/{\Phi_0}\right)$.  The resulting oscillation period is
${\Phi}_{0}$.

So far in our discussion we have neglected other types of charged excitations, particularly the charge $e/2$ excitations that are also expected at $\nu=5/2$. These excitations are always Abelian and their statistical angle is $\theta=\pi/2$, irrespective of the nature of the state. Eq.~(\ref{eq:net_phase_2}) then yields ${\Delta}{\gamma} = - 4{\pi}\left({\Phi}/{\Phi_0}\right)$ implying a ${\Phi}_{0}/2$ periodicity. While backscattering of these quasiparticles at quantum point contacts should be suppressed by comparison to $e/4$ quasiparticles (except in the anti-Pfaffian state), it is likely that their coherence length
is much longer, which in turn enhances their contribution to the coherent interference signal~\cite{Wan2008,Bishara2009a}. It therefore remains a puzzle that period ${\Phi}_{0}/2$
oscillations are not convincingly seen in our data, particularly in view of the fact that $e/2$ oscillations were a prominent feature seen in the side-gate voltage sweeps~\cite{Willett2009a,Willett2010a}. One possible explanation is the relatively narrow magnetic field window available for $B$-sweeps at $\nu=5/2$, which in turn makes it difficult to see longer-period oscillations. This window is limited by the narrow width of the $\nu=5/2$ plateau immediately flanked by the reentrant compressible integer states~\cite{Eisenstein2002,Xia2004} -- see more on this in the next section. It is also worth mentioning that the data presented here was measured at $T \approx 20$\,mK, while previously reported side-gate voltage sweeps were performed at $T\approx25$\,mK and higher; lowering the temperature should lead to the suppression of the $e/2$ contribution to oscillations. Testing this argument by measuring $B$-sweeps at higher temperatures and for different device sizes is an important future direction.

\subsection{Additional integer, 7/3, and 5/2 data sets; 5/2 splitting}
\label{sec:additional_data}

\begin{figure}[ht]
\centering
{\includegraphics[width=0.4\columnwidth]{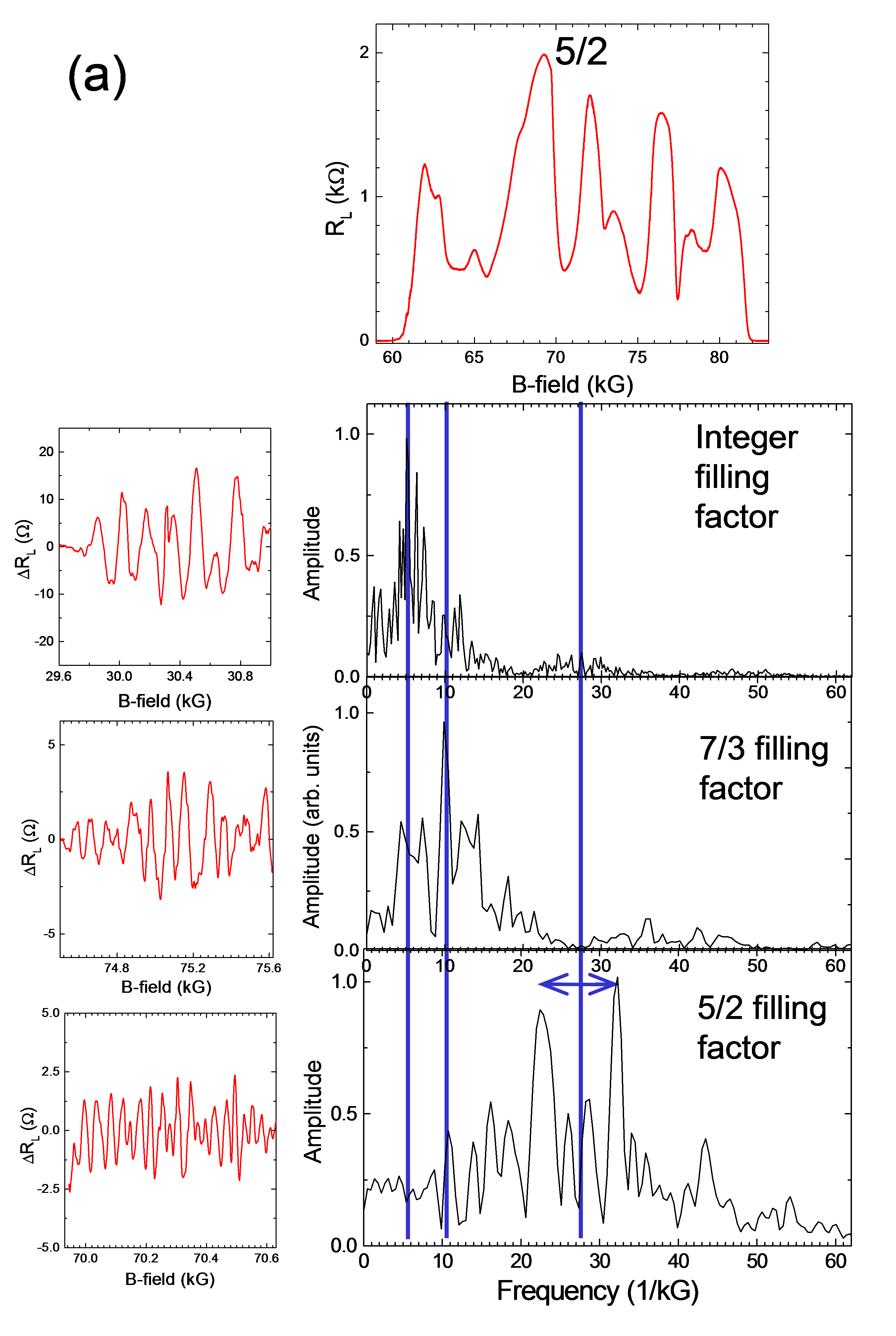}}
\hspace{1cm}
{\includegraphics[width=0.4\columnwidth]{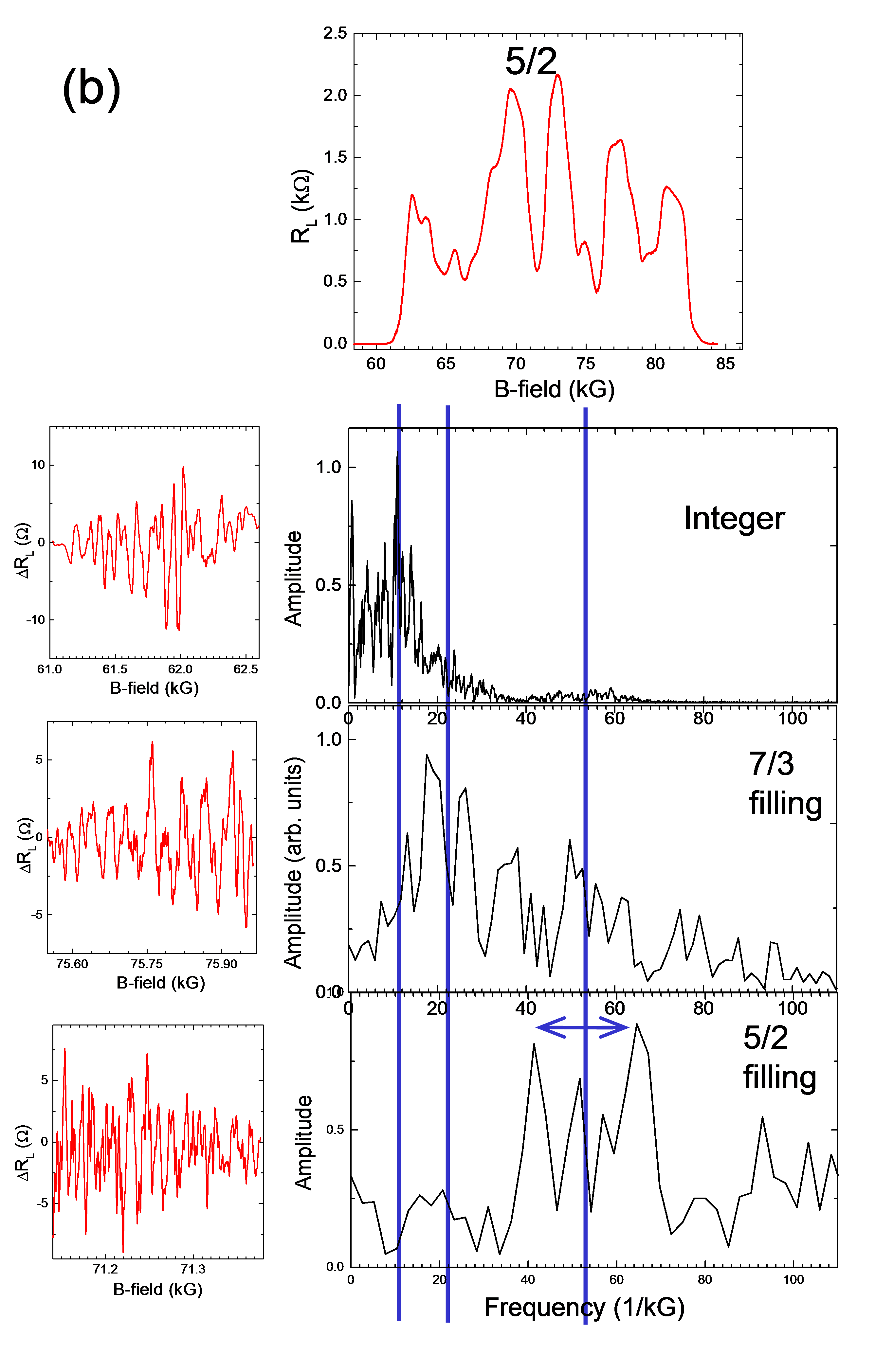}}
\newline
{\includegraphics[width=0.4\columnwidth]{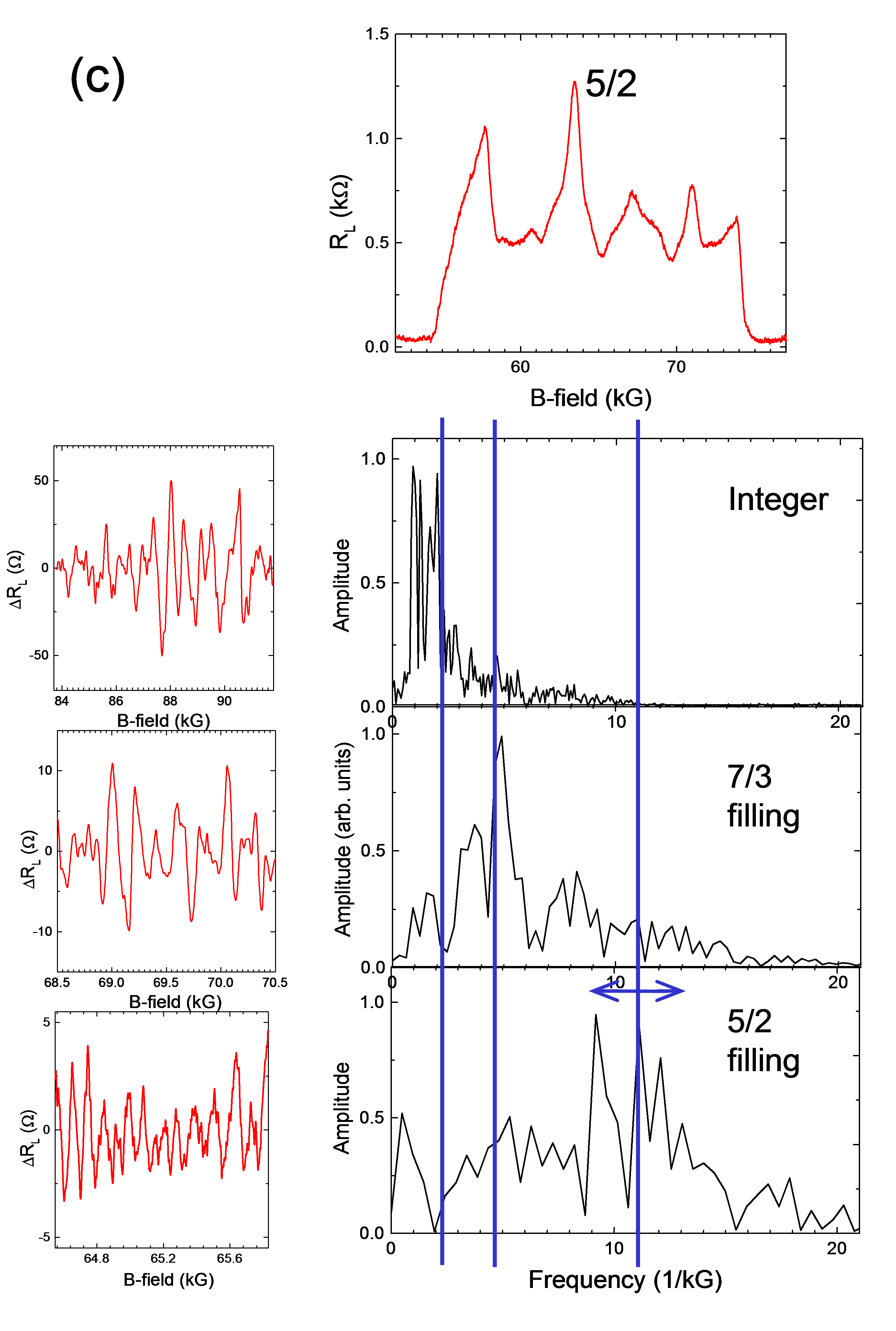}}
\hspace{1cm}
{\includegraphics[width=0.4\columnwidth]{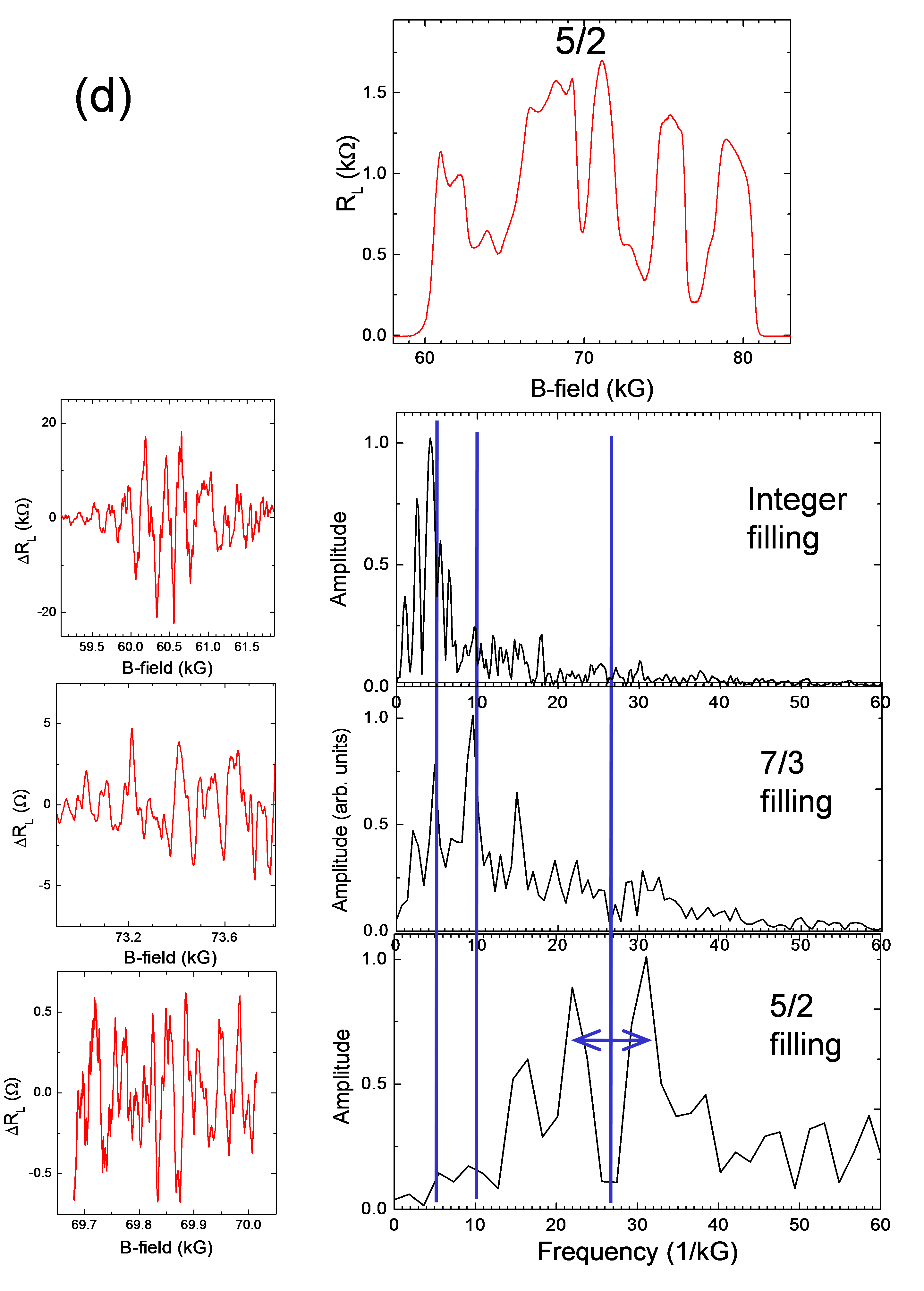}}
\caption{Four separate sample preparations showing transport
through the device (top panels), and $R\text{L}$ oscillations
(left panels) at integral, 7/3, and 5/2 filling factors. The right
hand panels show respective FFTs for those filling factors. The three
vertical blue lines mark the integer frequency ${1/\Phi_0}$, $2/{\Phi_0}$,
and $5/{\Phi_0}$. The 7/3 oscillation frequency is consistently at
twice that of the integer filling, and the 5/2 peak complex is centered
near 5 times the integer frequency. Data are taken at $T\approx 20$\,{m}K.}
\label{fig:S1}
\end{figure}

Additional sets of data showing Fourier transforms of oscillations at integer, $7/3$,
and $5/2$ filling factors from different sample preparations are shown in
Fig.~S\ref{fig:S1}. Note that the $5/2$ peak is split and is centered at
$5\Phi_0$ for all the data sets in this figure.

In roughly half the data sets examined, we were able to resolve the predicted
splitting of the 5/2 FFT peak at five times the fundamental integer
oscillation frequency ($1/{\Phi_0}$). This splitting at $5/{\Phi_0}$ to
$5/{\Phi_0}\pm 1/{\Phi_0}$, due to modulation of the non-Abelian
oscillation by the low frequency AB/anyonic oscillations at $\nu=5/2$,
can be seen only when we take the Fourier transform the resistance
over a substantial range in $B$-field: such data is displayed in Fig.~3 of the
main text, and in Fig.~S\ref{fig:S1}.

\begin{figure}
\centering
\includegraphics[width=0.78\columnwidth]{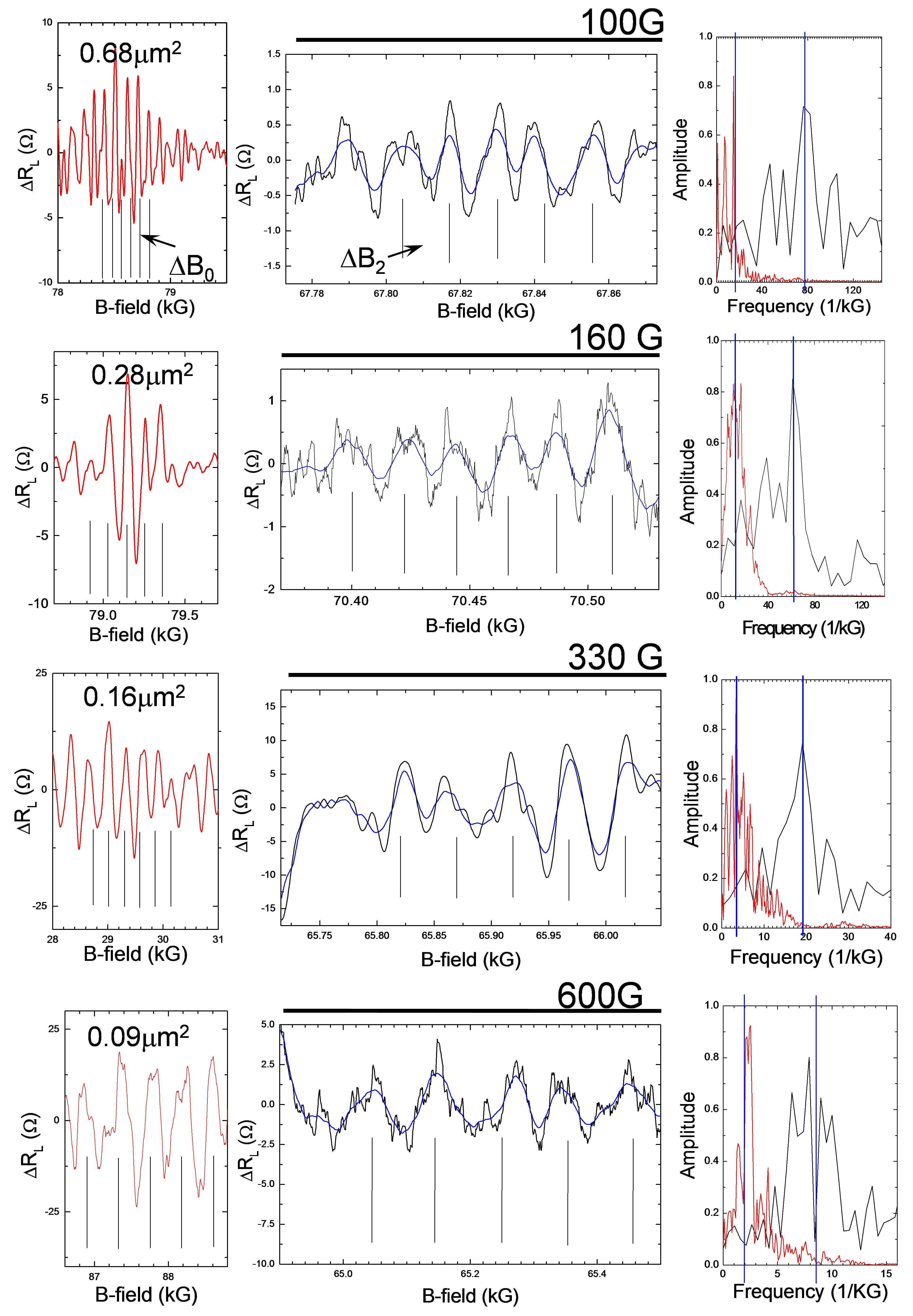}
\caption{AB oscillations at integer filling used to determine the area (left column) and the corresponding transport near
$\nu=5/2$ (center) showing small period oscillations consistent with
expression/suppression of non-Abelian $e/4$ interference, along with FFTs of
both spectra (right) for a series of interferometer
devices of different areas. In our model, the magnetic field ${\Delta}B_2$
necessary to add two $e/4$ quasiparticles is determined from
${\Delta}B_{2}/{\Delta}B_{0}=0.2$, or
${\Delta}B_{2}  A \approx 8G {\mu}\text{m}^{2}$. The
black line is an average of several (typically, eight) B-field sweeps, and the
blue line shows the same data smoothed by local averaging. In the FFT panel
the two vertical blue lines differ by a factor of 5 in frequency. Data are taken at $T\approx 20$\,{m}K.}
\label{fig:S2}
\end{figure}

Some $B$-field sweeps in this study ran over only a small range of $B$ near
5/2 filling to focus on measuring the small period (${\Phi_0}/5$)
oscillations attributed to non-Abelian $e/4$ expression/suppression:
these results are shown in Fig.~4 of the main text, and here in Fig.~S\ref{fig:S2}.
The other measurements, covering a larger range in $B$, noted above,
facilitate resolution of the splitting at 5/2. However, the range of
$B$-field of either of these oscillations is limited around 5/2 by the
presence of reentrant integer states~\cite{Eisenstein2002,Xia2004} adjacent
to $5/2$ filling on both the high- and low-field sides. This is a more prominent effect in the larger area devices.

\subsection{Gate sweep examples for multiple device sizes}
\label{sec:gate_sweeps}

\begin{figure}
\centering
\includegraphics[width=0.58\columnwidth]{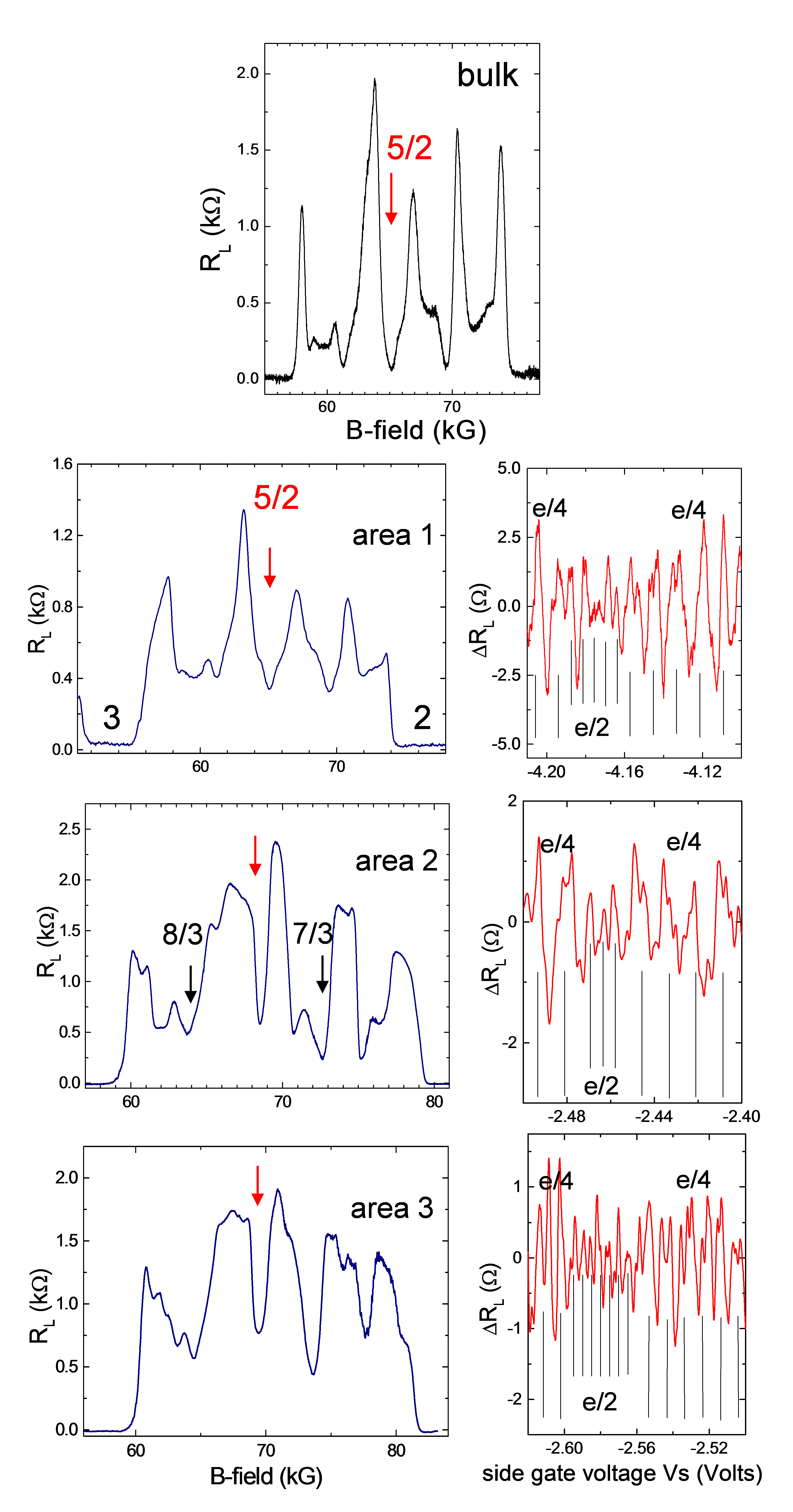}
\caption{Magneto-transport and gate-voltage sweep oscillations
at $\nu=5/2$.  All devices used here are fabricated from the same
heterostructure wafer, with representative bulk transport shown in the
left panel. The central column shows representative transport
through each device; note prominence of the $\nu=5/2$ minimum  and the presence of FQHE state at $\nu=7/3$ in these longitudinal resistance traces. The right hand column shows longitudinal resistance change with
side gate (2) sweep at $\nu=5/2$; each device demonstrates
oscillations consistent with the AB effect at periods corresponding to charge $e/4$.
The marked vertical lines of these periods are derived from similar
measurements at integers and $7/3$ fillings, defining the period corresponding to
$e/4$ charge.  In each device the previously observed~\cite{Willett2009a,Willett2010a} alternation
of $e/4$ and $e/2$ periods is affirmed in large side gate voltage
excursions. Temperature in all data is $T\approx 20$\,{m}K.}
\label{fig:S3}
\end{figure}

When side gate sweeps are applied rather than $B$-field sweeps,
the multiple devices of different areas examined here display the
same AB properties at 5/2 as observed in previous
studies~\cite{Willett2009a,Willett2010a}, namely alternation
of $e/4$ and $e/$2 period oscillations. See Fig.~S\ref{fig:S3}.  This alternation
is attributed to the side gate excursion changing not only the enclosed
magnetic flux number but also the localized non-Abelian $e/4$
quasiparticle number. There are oscillations corresponding to
AB interference of $e/4$ particles when an even number of $e/4$ quasiparticles
are enclosed, and the $e/2$ oscillations are apparent when that number is odd.
According to theoretical predictions \cite{Bishara2009a},
the $e/2$ oscillations are pervasive but are more easily observed when the larger amplitude
$e/4$ oscillations are suppressed.  The data of the Figure show this
alternation for all three rudimentary device sizes. Further
demonstration and details of these measurements can be found in
reference~\cite{Willett2013a}.

%\putbib[kirref]

\end{bibunit}

\clearpage

\onecolumngrid

\clearpage

\clearpage

\end{document}